\documentclass{amsart}
\usepackage{thmdefs}
\usepackage{graphicx}

\theoremstyle{definition}
\theoremstyle{remark}
\numberwithin{equation}{section}

\begin{document}
\title{Soft matrix models and Chern-Simons partition functions}
\author{M. Tierz}
\address{Instituto de Ciencias del Espacio (ICE/CSIC) \\
Institut d'Estudis Espacials de Catalunya (IEEC/CSIC) \\
Edifici Nexus, Gran Capit\`{a}, 2-4, 08034 Barcelona, Spain.}
\email{tierz@ieec.fcr.es}
\curraddr{}

\begin{abstract}
We study the properties of matrix models with soft confining potentials.
Their precise mathematical characterization is that their weight function is
not determined by its moments. We mainly rely on simple considerations based
on orthogonal polynomials and the moment problem. In addition, some of these
models are equivalent, by a simple mapping, to matrix models that appear in
Chern-Simons theory. The models can be solved with $q$ deformed orthogonal
polynomials (Stieltjes-Wigert polynomials), and the deformation parameter
turns out to be the usual $q$ parameter in Chern-Simons theory. In this way,
we give a matrix model computation of the Chern-Simons partition function on 
$S^{3}$ and show that there are infinitely many matrix models with this
partition function.
\end{abstract}

\maketitle

\bigskip

\smallskip

\section{Introduction}

Random matrix models \cite{Mehta,bretal} (\cite{zj} for a review) have
attracted great amount of interest both due to its mathematical structure and
for its manifold physical applications as well. When studying matrix models,
one is especially interested in the solution of matrix models in the large $%
N $ approximation, which corresponds to the planar Feynman diagrams \cite
{bretal,zj}. These models are characterized by the $N$ by $N$ matrix
variable $M$ and by an Hamiltonian $H=\mathrm{Tr}V\left( M\right) .$ After
diagonalization of the matrix, one is able to work in the eigenvalue space,
and to consider the following expression for the partition function: 
\begin{equation}
Z_{M}=\int \prod_{i=1}^{\lambda }d\lambda _{i}\exp \left( -V\left( \lambda
_{i}\right) \right) \prod_{i<j}\left( \lambda _{i}-\lambda _{j}\right)
^{\beta },  \label{part}
\end{equation}
with $\beta =1,2$ or $4$, depending on the symmetry. Equivalently, from the
perspective of random matrix theory one studies the joint probability
distribution function for the $N$ eigenvalues of the matrix \cite{Mehta}. It
has the well-known general form:

\begin{eqnarray}
P(x_{1},...,x_{N}) &=&C_{N}\prod_{i<j}\left| x_{i}-x_{j}\right| ^{\beta
}\prod_{i=1}^{N}\omega \left( x_{i}\right) =  \notag \\
&=&C_{N}\prod_{i<j}\left| x_{i}-x_{j}\right| ^{\beta }\exp
[-\sum_{i=1}^{N}V(x_{i})],
\end{eqnarray}
$\omega \left( x\right) $ and $V(x)$ are named weight function and confining
potential, respectively. If the elements of the random matrix are believed
to be statistically independent from each other, one obtains the quadratic
confinement potential $V\left( x\right) =x^{2}$, leading to the Gaussian
ensembles of random matrices \cite{Mehta}. In the usual physical
applications of matrix models \cite{zj}, the confining potentials $V\left(
x\right) $ are such that the weight function $\omega \left( x\right) =\exp
\left[ -V\left( x\right) \right] $ is determined by the knowledge of all of
its positive integer moments $\left\{ \gamma _{n}\right\} _{n=0}^{\infty },$
where $\gamma _{n}\equiv \int x^{n}\omega \left( x\right) dx.$ In this
paper, we want to study some general properties of matrix models with a
weight function that does not satisfy this property. Interestingly enough, a
very particular model in this category shows considerable interest in the
context of Chern-Simons theory \cite{Witt1,Labrev} and topological strings 
\cite{Witt2,GoVa2,Mar2}.

We briefly remind now the formalism of orthogonal polynomials. Recall that
the relevant quantities, such as the density of states and the correlation
functions, are obtained from the two-point kernel \cite{Mehta}:

\begin{equation}
K_{N}\left( x,y\right) =\sum_{k=0}^{N-1}\varphi _{k}\left( x\right) \varphi
_{k}(y),  \label{ker}
\end{equation}
with $\varphi _{k}\left( x\right) =\omega \left( x\right) p_{k}\left(
x\right) $, where $p_{k}\left( x\right) $ are the orthogonal polynomials
associated to the weight function $\omega \left( x\right) .$This is the
fundamental object, due to the following classical result \cite{Mehta}:

\begin{equation}
R_{k}(x_{1},x_{2},...,x_{k})=\det \left[ K_{N}\left( x_{p},x_{q}\right)
\right] _{q,p=1,2,...,n}.
\end{equation}
where the correlation functions are defined as:

\begin{equation}
R_{k}(x_{1},x_{2},...,x_{k})\equiv \frac{N!}{\left( N-k\right) !}%
\int_{-\infty }^{\infty }dx_{k+1}...\int_{-\infty }^{\infty }dx_{N}P\left(
x_{1},x_{2},...,x_{N}\right) .
\end{equation}
For Hermitian ensembles ($\beta =2$), the density of states is the kernel at
the origin:

\begin{eqnarray}
\rho (x) &=&\sum_{k=0}^{N-1}\varphi _{k}(x)^{2}  \label{dens} \\
&=&\omega \left( x\right) \left( p_{N}(x)p_{N-1}^{\prime }(x)-p_{N}^{\prime
}\left( x\right) p_{N-1}\left( x\right) \right) ,
\end{eqnarray}
the second expression follows when applying the Christoffel-Darboux formula 
\cite{Szego,Mehta}. For Gaussian ensembles for example, this quantity tends
to the well-known semi-circle law in the limit $N\rightarrow \infty $ \cite
{Mehta}. In this work we shall consider Hermitian ensembles, and this
expression for the density of states will be useful to understand relevant
conceptual points in our discussion.

As mentioned, we are concerned with the special properties of matrix models
with an indeterminate weight function. Our discussion is mainly based on the
classical moment problem \cite{Akh} (see \cite{sim} for a recent review,
that we will follow). The connection with random matrix theory can be
readily guessed by the fact that only the moments play a role in the
orthogonalization procedure. Therefore the two sets, the moments and the
orthogonal polynomials, are essentially equivalent.

The paper is organized as follows: In the next Section, we give a short
introduction to Chern-Simons theory and, focussing on these recent
developments, we show how the matrix models in \cite{Aga,Mar} are directly
related with an Hermitian matrix model with a log-normal weight function. In
this way, one can do exact computations through the associated orthogonal
polynomials. To illustrate this, we compute the partition function of
Chern-Simons theory on $S^{3}$ with gauge group $U(N).$ Interestingly
enough, the use of $q$-deformed orthogonal polynomials readily leads to the
natural parameter of Chern-Simons theory $q=\mathrm{e}^{-g_{s}}=\mathrm{e}^{%
\frac{2\pi i}{N+k}}$. Therefore, the orthogonal polynomials approach is a
method for non-perturbative solutions in Chern-Simons theory. In Section 3,
we show that there are actually infinitely many matrix models with the same
partition function. Finally, we show that this is a general feature of
models with very weak confining potential. To conclude, we present a summary
and possible directions for further research.

\section{Chern-Simons theory on $S^{3}$: Matrix Model formulation}

We begin by outlining basic and well-known facts about Chern-Simons theory,
emphasizing recent results and trends that connect Chern-Simons theory with
matrix models \cite{Aga,Mar}. As it is well-known, Chern-Simons theory is a
topological quantum field theory whose action is built out of a Chern-Simons
term involving as gauge field a gauge connection associated to a group $G$
on a three-manifold $M$. The action is: 
\begin{equation}
S(A)={\frac{k}{4\pi }}\int_{M}\mathrm{Tr}\left( A\wedge dA+{\frac{2}{3}}%
A\wedge A\wedge A\right) ,  \label{cs}
\end{equation}
with $k$ an integer number. The natural associated observables are the
correlators of Wilson loops, and its main interest come from the fact that
these correlators lead to quantum-group polynomial invariants of knots and
links \cite{Witt1}. For a review of the field since the seminal work \cite
{Witt1}, see \cite{Labrev}.

We mention now some few developments related to Chern-Simons theory. As
advanced in \cite{Witt2}, Chern-Simons gauge theory has a string description
in the sense of 't Hooft \cite{Hooft}. Indeed, as it is well-known, gauge
theories with the $SU(N)$ group admit a large $N$ expansion. In these
expansions, correlators are expanded in powers of $1/N$ while keeping the 't
Hooft coupling fixed $t=xN$, with $x$ the coupling constant of the gauge
theory. In the case of Chern-Simons theory, this large $N$ expansion is
reminiscent of a string theory expansion. This connection between
Chern-Simons and topological strings was first pointed out by Witten \cite
{Witt2} and has been extended in \cite{GoVa2}. For example, in \cite{Witt2}
it is shown that if one wraps $N$ $D$-branes on $M$ in $T^{*}M$, then the
associated topological A-model is a $U(N)$ Chern-Simons theory on the three
manifold $M$.

On the other hand, regarding matrix models, in \cite{GoVa} it was already
pointed out that the structure of the partition function of Chern-Simons
theory on $S^{3\text{ }}$ with gauge group $SU(N),$ resembles the usual
expression for the partition function of a one matrix model in terms of its
associated orthogonal polynomials. Moreover, in \cite{Mar}, it is shown that
the partition function of Chern-Simons theory on $S^{3}$ and with gauge
group $U(N)$, is given by:

\begin{equation}
Z=\frac{\mathrm{e}^{-\frac{g_{s}}{12}N\left( N^{2}-1\right) }}{N!}\int
\prod_{i<j}\left( 2\sinh \frac{u_{i}-u_{j}}{2}\right) ^{2}\prod_{i=1}^{N}%
\mathrm{e}^{-u_{i}^{2}/2g_{s}}\frac{du_{i}}{2\pi },  \label{sinh}
\end{equation}
that describes open topological $A$ strings on $T^{*}\emph{S}^{3}$ with $N$
branes wrapping $S^{3}$ (see the details in \cite{Mar}). These type of
models have been further considered in \cite{Aga,Halmagyi:2003fy}. In \cite{Diva}, it was
shown that topological strings for $B$-branes are equivalent to Hermitian
matrix models, then the idea in \cite{Aga} is to obtain the results in \cite
{Mar}, by applying mirror symmetry to obtain $B$-brane matrix models. We
will comment further on the results in \cite{Aga} later on. Now, we mainly
focus on the study of $\left( \ref{sinh}\right) $. For recent reviews along
these lines see \cite{Mar2}.

We remind that, as usual in Chern-Simons theory, the string coupling
constant $g_{s}$ is related with the $k$ in $\left( \ref{cs}\right) $ by: 
\begin{equation}
g_{s}=\frac{2\pi i}{k+N}.
\end{equation}
As explained in \cite{Mar}, the limit of the parameter $g_{s}\rightarrow 0,$
lead to the usual Gaussian Unitary ensemble. One can argue, for example,
that in this limit, the Gaussian becomes a Dirac delta function, and
therefore the results are independent of the correlation factor, or level
repulsion. In \cite{Aga,Mar}, this model is essentially studied through the
consideration of averages (in a Gaussian Unitary ensemble) of the following
quantities:

\begin{equation}
\mathrm{e}^{-\sum_{k=1}^{\infty }a_{k}\sigma _{k}\left( u\right) },\ \text{%
with }a_{k}=\frac{B_{2k}}{k\left( 2k\right) !}\ \text{and }\sigma _{k}\left(
u\right) =\sum_{i<j}\left( u_{i}-u_{j}\right) ^{2k}.
\end{equation}
This method leads to a great amount of perturbative information.
Nevertheless, we note that the connection with the usual Hermitian matrix
models -that is, of the type given by $(\ref{part})$ - is much simpler. Let
us consider the following simple change of variables: 
\begin{equation}
\mathrm{e}^{u_{i}}=x_{i},  \label{camb}
\end{equation}
then, $\left( \ref{sinh}\right) $, reads:

\begin{equation}
Z=\frac{\mathrm{e}^{-\frac{g_{s}}{12}N\left( N^{2}-1\right) }}{N!}\int
\prod_{i<j}\left( x_{i}-x_{j}\right) ^{2}\prod_{i=1}^{N}x_{i}^{-N}\mathrm{e}%
^{-\log ^{2}x_{i}/2g_{s}}\frac{dx_{i}}{2\pi }.  \label{lognorm}
\end{equation}

The factor $\prod_{i}x_{i}^{-(N-1)}$ can be readily absorbed by considering
a simple but remarkable property of the log-normal function $\omega \left(
x\right) =\mathrm{e}^{-\log ^{2}x_{i}/2g_{s}}:$

\begin{equation}
\omega \left( xq\right) =\sqrt{q}x\omega \left( x\right) ,  \label{func}
\end{equation}
Note that this functional equation is just a particular case of the
following elementary identity: 
\begin{eqnarray}
\mathrm{e}^{-k^{2}\log ^{2}\left( \mathrm{e}^{\left( -\alpha /2k^{2}\right)
}x\right) } &=&\mathrm{e}^{-k^{2}\left( \log ^{2}x+\alpha ^{2}/4k^{4}-\frac{%
\alpha }{k^{2}}\log x\right) }  \label{func2} \\
&=&\mathrm{e}^{-\alpha ^{2}/4k^{2}}x^{\alpha }\mathrm{e}^{-k^{2}\log ^{2}x}.
\notag
\end{eqnarray}

Therefore, the Chern-Simons matrix model is directly related to an ordinary
Hermitian ensemble with log-normal weight function $\omega \left( x\right) =%
\mathrm{e}^{-\log ^{2}x_{i}/2g_{s}}.$ Equivalently, one could have used the
change of variables $\exp \left( u_{i}+Ng_{s}\right) =x_{i}$. Notice that
the term $Ng_{s}$ is necessary to cancel, together with the term coming from
the Jacobian, the contribution $\prod_{i}x_{i}^{-(N-1)}=\prod_{i<j}\frac{1}{%
x_{i}x_{j}}.$ Actually, this method was already employed in \cite{For},
where the same matrix model seems to appear in a very different context. One
of our goals is to make contact between this model and Chern-Simons theory.
As we shall see, this will comes out rather easily, due to the existence of
a closed system of orthogonal polynomials associated to the log-normal
weight. These polynomials are $q$-deformed polynomials known as
Stieltjes-Wigert polynomials \cite{Szego}. This opens the possibility of a
complete solution for the model $\left( \ref{sinh}\right) $ of \cite
{Aga,Mar} (see also \cite{For}), using the classical results for Hermitian
ensembles and these polynomials. Secondly, we shall show -giving explicit
examples- that there exists infinitely many matrix models -deformations of $\left( \ref{lognorm}\right) $-
with exactly the same (Chern-Simons) partition function.

\subsection{Chern-Simons partition function through the matrix model
computation}

Using the previous results and the Stieltjes-Wigert orthogonal polynomials 
\cite{Szego}, we can find a complete solution of the matrix model considered
in \cite{Mar}, and, in particular, for the partition function. For this, we
can use the following well-known result for the partition function in terms
of the orthogonal polynomials \cite{zj}: 
\begin{equation}
Z=\int ...\int \prod_{i=1}^{N}\omega \left( x_{i}\right)
dx_{i}\prod_{i<l}\left( x_{i}-x_{l}\right) ^{2}=\frac{N!}{%
\prod_{i=0}^{N-1}a_{i}^{2}}=N!a_{0}^{-2N}\prod_{i=1}^{N-1}\left( \left( 
\frac{a_{i-1}}{a_{i}}\right) ^{2}\right) ^{N-i}.
\end{equation}
where the coefficient $a_{i}$ is: 
\begin{equation}
p_{i}\left( x\right) =a_{i}x^{i}+...
\end{equation}

The first step is to compute the $Z$ associated to the Stieltjes-Wigert
orthogonal polynomials. The coefficients are given by \cite{Szego}:

\begin{equation}
a_{j}=q^{\left( j+1/2\right) ^{2}}\left\{ \left( 1-q\right) ...\left(
1-q^{j}\right) \right\} ^{-1/2},
\end{equation}
then 
\begin{equation}
\left( \frac{a_{j-1}}{a_{j}}\right) ^{2}=q^{-4j}\left( 1-q^{j}\right) ,
\end{equation}
and $a_{0}=q^{1/4}.$ Therefore: 
\begin{eqnarray}
Z_{sw} &=&N!q^{-N/2}\prod_{j=1}^{N-1}q^{-4j\left( N-j\right) }\left(
1-q^{j}\right) ^{N-j} \\
&=&\allowbreak N!\allowbreak q^{-\frac{1}{6}N\left( 2N-1\right) \left(
2N+1\right) }\prod_{j=1}^{N-1}\left( 1-q^{j}\right) ^{N-j},  \notag
\end{eqnarray}
but we are interested in :

\begin{eqnarray}
&&Z_{\sinh }=\int \prod_{i=1}^{N}\frac{du_{i}}{2\pi }\mathrm{e}^{-\frac{%
u_{i}^{2}}{2g_{s}}}\prod_{i<j}\left( 2\sinh \left( \frac{u_{i}-u_{j}}{2}%
\right) \right) ^{2} \\
&=&\left( 2\pi \right) ^{-N}\mathrm{e}^{\frac{-N^{3}g_{s}}{2}}\int
\prod_{i=1}^{N}dx_{i}\mathrm{e}^{-\frac{\log ^{2}\left( x_{i}\right) }{2g_{s}%
}}\prod_{i<j}\left( x_{i}-x_{j}\right) ^{2}.  \notag
\end{eqnarray}
Making the identification $k^{2}=\frac{1}{2g_{s}}$ , and therefore $q=%
\mathrm{e}^{-g_{s}}$. It is remarkable that we have naturally obtained, from
this $q$ deformed orthogonal polynomials, the usual $q$ parameter in
Chern-Simons theory. The partition function with the usual Stieltjes-Wigert
weight, corresponds to: 
\begin{eqnarray}
Z_{sw} &=&\int \prod_{i=1}^{N}dx_{i}\pi ^{-1/2}k\mathrm{e}^{-k^{2}\log
^{2}\left( x_{i}\right) }\prod_{i<j}\left( x_{i}-x_{j}\right) ^{2} \\
&=&\left( 2\pi g_{s}\right) ^{-N/2}\int \prod_{i=1}^{N}\mathrm{e}^{-\frac{%
\log ^{2}\left( x_{i}\right) }{2g_{s}}}dx_{i}\prod_{i<j}\left(
x_{i}-x_{j}\right) ^{2}.  \notag
\end{eqnarray}
Therefore: 
\begin{equation}
Z_{\sinh }=\left( \frac{g_{s}}{2\pi }\right) ^{N/2}N!\mathrm{e}^{\frac{1}{6}%
g_{s}N\left( N^{2}-1\right) }\prod_{j=1}^{N-1}\left( 1-q^{j}\right) ^{N-j}.
\end{equation}
To make explicit contact with the typical expressions for the Chern-Simons
partition functions (as in \cite{GoVa} for example) we make some
transformations on the product term.

\begin{eqnarray}
\prod_{j=1}^{N-1}\left( 1-q^{j}\right) ^{N-j} &=&\prod_{j=1}^{N-1}\left(
2\sinh \frac{g_{s}j}{2}\right) ^{N-j}\mathrm{e}^{-\frac{g_{s}j\left(
N-j\right) }{2}}  \notag \\
&=&\mathrm{e}^{\frac{-1}{12}g_{s}N\left( N^{2}-1\right) }\mathrm{e}^{\frac{%
i\pi }{4}N\left( N-1\right) }\cdot \prod_{j=1}^{N-1}\left( 2\sin \frac{g_{s}j%
}{2i}\right) ^{N-j}.
\end{eqnarray}
The final expression for the partition function is \cite{Mar}: 
\begin{eqnarray}
Z &=&\frac{\mathrm{e}^{-\frac{1}{12}g_{s}N\left( N^{2}-1\right) }}{N!}%
Z_{\sinh } \\
&=&\left( \frac{g_{s}}{2\pi }\right) ^{N/2}\mathrm{e}^{\frac{i\pi }{4}%
N\left( N-1\right) }\prod_{j=1}^{N-1}\left( 2\sin \frac{g_{s}j}{2i}\right)
^{N-j},  \notag
\end{eqnarray}
and since $g_{s}=\frac{2\pi i}{k+N}$, we finally find: 
\begin{equation}
Z=\mathrm{e}^{\frac{1}{4}i\pi N^{2}}\left( k+N\right)
^{-N/2}\prod_{j=1}^{N-1}\left( 2\sin \frac{\pi j}{k+N}\right) ^{N-j}.
\end{equation}

It was already mentioned in \cite{GoVa}, that the structure of the partition
function resembled very much the general expression for the partition
function of a one matrix model in terms of the associated orthogonal
polynomials. Here, based on the result in \cite{Mar}, we are making this
statement precise through the explicit computation. Notice that the
differences between the $SU(N)$ and $U(N)$ comes essentially from the
contribution of the partition function of $U(1)$ (see, for example, the
first reference in \cite{Mar2} for more details).

\section{Infinitely many matrix models with the same Chern-Simons partition
function}

In this second part we shall show that the matrix models involved that
appear in Chern-Simons theory present a novelty in comparison with most of
known matrix models (like the ones that are relevant in 2D quantum gravity).
Broadly speaking, its weight function is so broad that there is a
non-uniqueness issue that leads to very practical and concrete results for
the matrix model: there are infinitely many models with identical
properties. In particular, identical (Chern-Simons) partition function. We
begin by briefly reviewing some introductory results in the theory of the
moment problem.

\subsection{Relevant results from the moment problem.}

The two basic moment problems are \cite{sim}:

\subsubsection{Hamburger moment problem.}

Given a sequence of reals $\gamma _{0,}\gamma _{1},...$ , when is there a
measure, $d\rho ,$ on $\left( -\infty ,\infty \right) $ so that 
\begin{equation}
\gamma _{n}=\int_{-\infty }^{\infty }x^{n}d\rho \left( x\right)
\end{equation}
and if such measure exists, is it unique ?

\subsubsection{Stieltjes moment problem.}

Given a sequence of reals $\gamma _{0,}\gamma _{1},...$ , when is there a
measure, $d\rho ,$ on $\left( 0,\infty \right) $ so that 
\begin{equation}
\gamma _{n}=\int_{0}^{\infty }x^{n}d\rho \left( x\right)
\end{equation}
and if such measure exists, is it unique ?

In our case, we are not concerned with the question regarding existence,
since we always refer to the weight function $\omega \left( x\right) $ and
its associated moments. Then, the main point is to know whether we are in
the determinate or indeterminate case. To fix ideas, let us mention that
there is a natural boundary for functions of the form $\mathrm{e}^{-V(x)},$
with $V(x)$ of polynomial form. In the Hamburger case, the boundary is given
by $V(x)=\left| x\right| ^{\alpha }$ with $\alpha =1$, while in the
Stieltjes case it is given by $V(x)=\left| x\right| ^{\alpha }$ with $\alpha
=1/2$.

Then, in the usual matrix models considered (as reviewed in \cite{zj} for
example), one is essentially always in the determined case.

Consider now the following generalization of the log-normal weight:

\begin{equation}
f_{\theta }(x)=\mathrm{e}^{-\log ^{2}x}(1+\vartheta \sin \left( 2\pi \log
x)\right) ,\quad \text{with}\ \vartheta \in \left[ -1,1\right] .  \label{sti}
\end{equation}
The moments of this function are \cite{Sti}: 
\begin{equation}
\gamma _{n}=\int_{0}^{\infty }f_{\theta }(x)x^{n}dx=\sqrt{\pi }\mathrm{e}^{%
\frac{1}{4}(n^{2}+1)}.  \label{mom}
\end{equation}
Note that all the integer moments are completely independent of the
parameter $\vartheta $. This means that all the functions in the family $%
f_{\theta }(x)$ have the same moments. Thus, they are all undetermined by
them. Conversely, one can say that the set of moments $\gamma _{n}=\sqrt{\pi 
}\mathrm{e}^{\frac{1}{4}(n^{2}+1)},$ is an indeterminate set. We present now
a detailed consideration of this family of functions. The main point
is the computation of its Mellin transform.

First, note that for $\theta =0$ we have the log-normal distribution $%
\mathrm{e}^{-\log ^{2}\left( x\right) }.$ Its Mellin transform is: 
\begin{equation}
M\left[ f\left( x\right) ;s\right] =\mathrm{e}^{\pi ^{2}/4}\mathrm{Erf}%
\left( -\frac{s}{2}\right) .
\end{equation}
The other part of the Stieltjes function is $\mathrm{e}^{-\log ^{2}x}\sin
\left( 2\pi \ln x\right) $. From $(\ref{mom}),$ we know that all of its
integer moments are zero. The most enlightening possibility is to compute
its Mellin transform:

\begin{figure}[tph]
\centering
\includegraphics{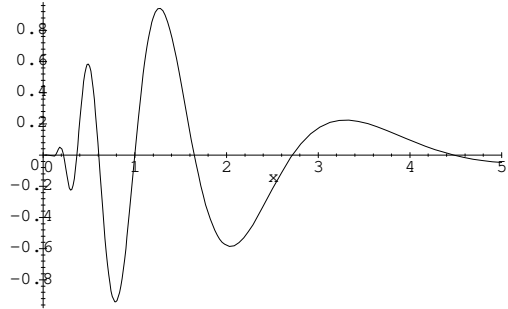}
\caption{The oscillatory term in $f_{\theta }(x)$}
\end{figure}
\begin{equation}
I\equiv \int_{0}^{\infty }x^{s-1}\mathrm{e}^{-\log ^{2}\left( x\right) }\sin
\left( 2\pi \log x\right) dx.
\end{equation}
We use the change of variables $y=-\frac{s+1}{2}+\log x.$ The integral
becomes: 
\begin{eqnarray}
I &=&\int_{-\infty }^{\infty }\mathrm{e}^{-y^{2}+\frac{1}{4}\left(
s^{2}+1\right) }\sin \left[ 2\pi \left( y+\frac{s+1}{2}\right) \right] dy \\
&=&\sqrt{\pi }\mathrm{e}^{\frac{1}{4}\left( s^{2}+1\right) }\sin \left( \pi
\left( 1+s\right) \right) .  \notag
\end{eqnarray}

The change of variables is the one employed by Stieltjes \cite{Sti}, with a generic,
complex $s$ instead of an integer $k$. Clearly, after the change of
variables, since $\sin $ is a periodic and odd function, its value is zero
at $s=k.$ Nevertheless, the result is even clearer when looking at the full
Mellin Transform: the contribution of the additional function is zero only
for integer values of $s.$

As a consequence, for example, we have that any of the following functions
have the same integer moments.

\begin{figure}[htp]
\centering
\includegraphics{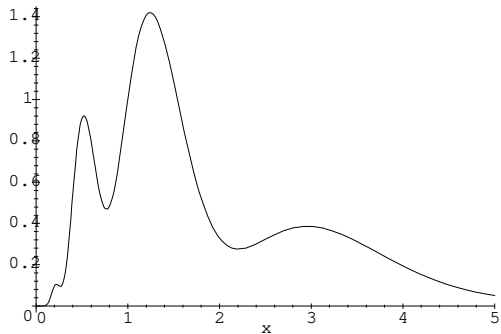}
\caption{$\theta=1/2$}
\end{figure}

\begin{figure}[htp]
\centering
\includegraphics{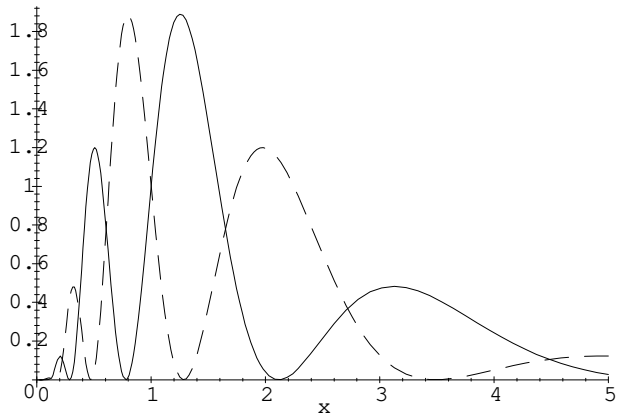}
\caption{$\theta=1$ (solid) and $\theta=-1$ (dashed)}
\end{figure}

Consequently, it is manifest now that there are infinitely many matrix
models with exactly the same partition function. Therefore, it turns out
that the partition function of $U\left( N\right) $ Chern-Simons theory on $%
S^{3}$ can be given by an infinitely many different matrix models. From what
we have seen, one possibility is:

\begin{equation}
Z_{N}=\int \prod_{i=1}^{N}\mathrm{e}^{-u_{i}^{2}/2g_{s}}\left( 1+\theta \sin
\left( 2\pi \left( u_{i}/g_{s}+N\right) \right) \right) \frac{du_{i}}{2\pi }%
\prod_{i<j}\left( 2\sinh \frac{u_{i}-u_{j}}{2}\right) ^{2},  \label{gen}
\end{equation}
as always, with $\theta \in \left[ -1,1\right] .$ Notice that due to the
correspondence between $(\ref{sinh})$ and $(\ref{lognorm})$ through $(\ref
{camb})$, the density of states of $(\ref{gen})$ does \emph{not} satisfy the
property: 
\begin{equation}
\int_{-\infty }^{\infty }\rho _{\theta }\left( u\right) u^{n}du=\gamma _{n},
\end{equation}
(that is, the moments are different for each matrix model). But rather: 
\begin{equation}
\int_{-\infty }^{\infty }\rho _{\theta }\left( \mathrm{e}^{u}\right) \mathrm{%
e}^{\left( n+1\right) \left( u+Ng_{s}\right) }du=\widetilde{\gamma }_{n},\
with\ n\in Z.  \label{expmom}
\end{equation}

To extract further physical consequences of these results and its
implications for the connection between matrix models and topological
quantum field theories seems an interesting open question.

\section{On matrix models with soft potentials}

In this concluding Section we show that the properties exhibited by the
Chern-Simons matrix model are typical of models with weakly confining -%
\textit{soft}- potentials. The mathematically precise description is that
the potential is such that $\mathrm{e}^{-V\left( x\right) }$ is a function
that is not determined by its moments. This feature has already been
discussed in the context of random matrix theory and also in the particular
case of $q$-deformed matrix models \cite{Mutt}. In this last case, the one
we are mainly interested, the non-uniqueness is strong since the potentials
are asymptotically $V(x)\sim \log ^{2}x$ for $x\rightarrow \infty $ \cite
{Mutt} and we have already mentioned that we are in the soft regime when the
confinement provided by the potential is weaker than $V(x)\sim x$.

Employing basic results on indeterminate functions $w\left( x\right) ,$ and
the classical random matrix formulae, one can prove that the density of
states of a matrix model is an indeterminate function if its weight function
is indeterminate by its moments. This can be readily shown if we consider
the expression for the density of states (for any $N$) $\left( \ref{dens}%
\right) $ and Krein proposition \cite{sim}, that essentially says that if $%
\int_{-\infty }^{\infty }\frac{\log Q_{N}(x)}{1+x^{2}}dx$ is convergent,
then the corresponding moment problem is undetermined. An analogous case
holds for the Stieltjes case.

From $\left( \ref{dens}\right) $ we have: 
\begin{eqnarray}
\log \rho (x) &=&\log \omega \left( x\right) +\log \left(
p_{N}(x)p_{N-1}^{\prime }(x)-p_{N}^{\prime }\left( x\right) p_{N-1}\left(
x\right) \right)  \notag \\
&=&\log \omega \left( x\right) +\log Q_{N}(x).  \label{2}
\end{eqnarray}
with $Q_{N}(x)$ a polynomial. We write the expression in the second way, to
emphasize that we just have the addition of a polynomial term. Then: 
\begin{equation}
\int_{-\infty }^{\infty }\frac{\log Q_{N}(x)}{1+x^{2}}dx<\infty ,
\end{equation}
and 
\begin{equation}
\int_{0}^{\infty }\frac{\log Q_{N}(x)}{\sqrt{x}\left( 1+x\right) }dx<\infty .
\end{equation}
Therefore, the convergence or divergence of $\int_{-\infty }^{\infty }\frac{%
\log \rho (x)}{1+x^{2}}dx$ and $\int_{0}^{\infty }\frac{\log \rho (x)}{\sqrt{%
x}\left( 1+x\right) }dx$ is exclusively given by the convergence or
divergence of $\int_{-\infty }^{\infty }\frac{\log \omega (x)}{1+x^{2}}dx$
and $\int_{0}^{\infty }\frac{\log \omega (x)}{\sqrt{x}\left( 1+x\right) }dx,$
respectively. Then, taking into account Krein proposition, it turns out that
the matrix models characterized by an indeterminate weight function, have
different (but indeterminate) density of states but the same partition
function. Thus, we have seen that the density of states is an indeterminate
function. Let us show explicitly that their moments are all equal: 
\begin{equation}
\rho _{\theta }(x)=\omega _{\theta }\left( x\right) Q_{N}\left( x\right) ,
\end{equation}
that is clearly $\theta $ dependent and the moments: 
\begin{equation}
\delta _{n},_{\theta }=\int x^{n}\omega _{\theta }\left( x\right)
Q_{N}\left( x\right) dx=\delta _{n}.
\end{equation}
are $\theta -$independent, since the polynomial $Q_{N}\left( x\right) $ is
always the same and then we are lead to a sum of moments of the
indeterminate function $\omega _{\theta }\left( x\right) $, each one of them 
$\theta -$independent. Now, we can show that the partition functions are
identical. We use the following expression: 
\begin{equation}
Z_{N,\theta }=\int \rho _{N}\left( x\right) dx=\int \omega _{\theta }\left(
x\right) Q_{N}(x)dx=Z_{N}.
\end{equation}
That is, the partition function as the $0$-moment of the density of states.
Alternatively, the usual expression for the partition function in terms of
the orthogonal polynomials can be used \cite{zj}: 
\begin{equation}
Z_{N}=\frac{N!}{\prod_{i=0}^{N-1}a_{i}^{2}},  \label{partfunc}
\end{equation}
where $a_{i}$ are the following coefficients of the associated orthogonal
polynomials: $P_{N}\left( x\right) =a_{N}x^{N}+...$ This expression is valid
for any weight function and only depends on the associated orthogonal
polynomials, that are the same for any $\theta .$

Then, for example, all matrix models with the weight functions $(\ref{sti})$%
, have the same orthogonal polynomials (essentially the Stieltjes-Wigert 
\cite{Szego}, see above). Nevertheless, it is plain that each one will have
a different density of states, since we have a different weight function (as
we have seen in the previous figures).

In any case, it is clear that approaching matrix models only from the
perspective of orthogonal polynomials, there is no a priori reason to choose
the log-normal weight, since any member of the infinite family has identical
orthogonal polynomials. Furthermore, the family of functions above discussed
is by no means the only one. From our analysis with the Mellin transform, is
clear that many other functions instead of $\sin $, can lead to the same
phenomena. In addition, examples such as \cite{Ask}: 
\begin{equation}
\omega \left( x\right) =\frac{x^{-5/2}}{\left( -x;q\right) _{\infty }\left(
-q/x;q\right) _{\infty }},\quad 0<x<\infty ,
\end{equation}
can be relatively easily found in the literature. This weight function also
leads to the Stieltjes-Wigert polynomials, that are actually terminating
basic hypergeometric series and have intimate relationships with theta
functions \cite{Ask}. In this work, we have discovered their remarkable
connection with Chern-Simons theory and quantum topological invariants. In
general, it turns out that the set $V$ of solutions to an indeterminate
moment problem contains very different types of measures: measures $\mu \in
V $ with a $C^{\infty }-$density, discrete measures and measures which are
continuous singular \cite{ber}. Explicit cases of these last types are found
in \cite{Chi}.

\section{ Conclusions and outlook}

\medskip We have seen that the matrix models that appear in Chern-Simons
theory are exactly solvable and that have many interesting properties, which
are novel in comparison with the matrix models that appear in 2D quantum
gravity. Therefore, the orthogonal polynomials technique constitute an
interesting non-perturbative method. We have checked this for the simple
case of $S^{3}$ and gauge group $U(N)$ and, interestingly enough, the use of 
$q$-deformed orthogonal polynomials readily leads to the usual parameter of
Chern-Simons theory. It seems feasible, and certainly an interesting task,
to try to extend this result to other geometries -like lens spaces- and to
other gauge groups as well \cite{Halmagyi:2003fy}.

On the other hand, the non-uniqueness property that we have established does
not only seem to be interesting from a mathematical perspective, but it may
have some physical meaning as well. We have emphasized the fact that the
original Chern-Simons matrix model can be deformed in infinitely many ways,
leaving the partition function always unmodified. Since this is a
Chern-Simons partition function, we know it is a topological invariant. A
well-known result in topology says that the topological invariant of a
three-manifold is independent of the triangulation employed to compute the
invariant. Thus, one is tempted to try to interpret all these matrix models
as representing different triangulations of the manifold. Thus, in a sense,
it would be interesting to try to obtain an analogous interpretation to the
role of more ordinary matrix models in 2D quantum gravity, but in three
dimensions. This is a bit speculative line of research, but it may still be
worth. In any case, the non-uniqueness properties are a distinctive
mathematical feature of the models and the way the it appears seems also
rather interesting in itself, since the resulting logarithmic oscillations
are quite typical of physical models where a quantum group structure is
present. A more detailed account of this will appear elsewhere.

\bigskip

\textbf{Acknowledgments}

\medskip

The author is indebted to Marcos Mari\~{n}o for very enlightening
discussions and correspondence, and also for a careful reading of the paper.

\end{document}